\documentclass[twocolumn,prl,amsmath,amssymb]{revtex4}
\usepackage{graphicx}
\usepackage{dcolumn}
\usepackage{bm}

\begin{document}

\title{Membrane morphology induced by anisotropic proteins}
\author{Kiyotaka Akabori}
\email{kakabori@physics.umass.edu}
\author{Christian D. Santangelo}
\email{csantang@physics.umass.edu}
\affiliation{
Department of Physics, University of Massachusetts, Amherst, MA  01003}
\date{\today}

\begin{abstract}
There are a great many proteins that localize to and collectively generate curvature in biological fluid membranes. We study changes in the topology of fluid membranes due to the presence of highly anisotropic, curvature-inducing proteins. Generically, we find a surprisingly rich phase diagram with phases of both positive and negative Gaussian curvature. As a concrete example modeled on experiments, we find that a lamellar phase in a negative Gaussian curvature regime exhibits a propensity to form screw dislocations of definite burgers scalar but of both chirality. The induced curvature depends strongly on the membrane rigidity, suggesting membrane composition can be a factor regulating membrane sculpting to to curvature-inducing proteins.
\end{abstract}


\maketitle

To form the structures of many cellular organelles from the Golgi apparatus to the endoplasmic reticulum to the mitochondrial inner membrane, lipid bilayers must be molded and shaped by an army of proteins \cite{nature}. Among these are proteins known to localize to and, collectively, induce curvature in fluid membranes \cite{nature, BAR, bax}. A flurry of theory \cite{fournier, iglic, iglic2, may}, simulations \cite{deserno, voth}, and experiments \cite{wong, wong2} has established that proteins can, in principal, induce large-scale shape changes in membranes. Nevertheless, how these proteins control morphology and what other factors are important in determining morphology is still poorly understood.

In this paper, we develop a mean-field model to describe how anisotropic, curvature-inducing proteins induce topological changes in membranes. Using our general model, we show that the magnitude and sign of the induced membrane curvature depends on the protein intrinsic curvatures and concentrations as well as on the rigidities of the underlying membrane. Since the effective membrane rigidity can depend on lipid composition, our model qualitatively explains the observed role of membrane composition in determining curvature \cite{wong,wong2}. For concreteness, we explicitly apply our model to the formation of bicontinuous structures in a lamellar phase doped with saddle-forming proteins. Note only have bicontinuous, cubic phases been observed in model membranes using a variety of proteins \cite{wong2, wong3}, but they are also seen \textit{in vitro} in the inner membranes of the mitochondria of starved ameoba \cite{dsurface}.

\begin{figure}[b]
\includegraphics[width=3.5in]{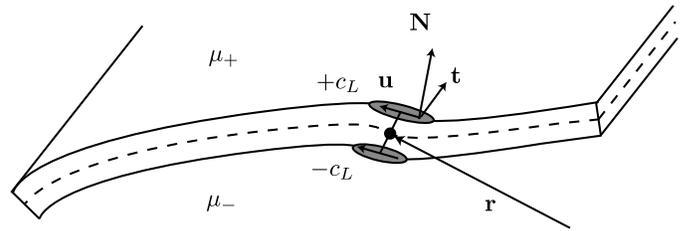}
\caption{\label{fig:fig1} Schematic model of proteins binding to a bilayer membrane. The proteins are at the same position, $\mathbf{r}$, under a normal projection to the midsurface (dashed) with $\mathbf{u}$ and $\mathbf{t}$ tangent to the membrane. If proteins bound to the top layer induce a local curvature $+c_L$ along the midsurface, an identical protein bound to the bottom surface induces a curvature $-c_L$.}
\end{figure}
To each bound protein, we associate a position $\mathbf{r}$ on a two-dimensional surface by projecting the protein center along the normal to the midsurface. A unit vector $\mathbf{u}$, tangent to the surface, points along its backbone  (see Fig. \ref{fig:fig1}). The second fundamental form of the midsurface can be expressed in terms of two principal curvatures $c_1$ and $c_2$ as $h_{i j} = c_1 \mathbf{e}_{1,i} \mathbf{e}_{1,j} + c_2 \mathbf{e}_{2,i} \mathbf{e}_{2,j}$, where the $\mathbf{e}_i$ are unit vectors along the principal curvature directions. Finally, we define a unit vector $\mathbf{t} = \mathbf{N} \times \mathbf{u}$, where $\mathbf{N}$ is the surface normal, that points transverse to the protein backbone. It will prove convenient to define the angle $\cos \theta = \mathbf{u} \cdot \mathbf{e}_1$, where the dot product is taken with respect to the surface metric, and dimensionless prescribed curvatures $\xi_i = c_i/|c_L|$. Assuming Hookean elasticity for the protein-membrane interaction, we obtain the energy
\begin{eqnarray}
\frac{E_\pm(\mathbf{r},\theta)}{k_B T} &=& \frac{\lambda_L}{2} \left(\xi_1 \cos^2 \theta + \xi_2 \sin^2 \theta \mp 1 \right)^2\nonumber\\  \label{eq:Ep}
& & + \lambda_X \left[\left(\xi_1-\xi_2\right) \cos \theta \sin \theta \mp \xi_X\right]^2\\
& & + \frac{\lambda_T}{2} \left( \xi_1 \sin^2 \theta + \xi_2 \cos^2 \theta \mp \xi_T\right)^2,\nonumber
\end{eqnarray}
where the sign refers to the sign of $\xi_L$. The first term describes the interaction along $\textbf{u}$ while the last term describes the transverse interaction, which induces the curvature $\xi_T c_L = c_T$ along $\mathbf{t}$. One can think of the $\lambda_i$ as the energetic cost, in units of $k_B T/2$, of binding an intrinsically-curved protein to a flat membrane. When $\lambda_T=\lambda_L=\lambda_X$ do we recover the previously studied models \cite{fournier, iglic}.

Rather than fixing the protein density, we calculate in a fixed chemical potential ensemble. Since the proteins on each leaf need not be in equilibrium, we must introduce different chemical potentials, $\mu_\pm$. If a protein, when bound to one leaf of a bilayer, induces a curvature $\xi_i$, it  must induce a curvature $-\xi_i$ when bound to the opposite leaf (Fig. \ref{fig:fig1}). Therefore, we associate $\mu_\pm$ with the positive or negative sign of Eq. (\ref{eq:Ep}) respectively. Since the proteins only interact through the membrane curvature, the partition function for either membrane leaf is found simply by exponentiating the single protein partition function,
\begin{eqnarray}
\Omega_\pm
&=& \exp \left\{ \int \frac{dA~d\theta}{A_0} z_\pm e^{-E_\pm(\theta,\mathbf{r})/(k_B T)} \right\},\label{eq:Z}
\end{eqnarray}
where $dA$ is the area measure on the membrane midsurface, $A_0$ is a characteristic cross-sectional area of the protein and $z_\pm = \exp[mu_\pm/(k_B T)]$ are the protein fugacities. The average protein concentration and orientation is, therefore, $\langle \rho(\mathbf{r},\theta) \rangle=z_\pm \exp \{ -E_\pm(\mathbf{r},\theta)/(k_B T) \}/(2 \pi A_0)$.

Finally, we combine Eq. (\ref{eq:Z}) with the Helfrich energy for a membrane, finding
\begin{eqnarray}
\frac{F}{k_B T} &=& 
\label{eq:energy}
\int dA~\left[ \frac{\kappa}{2} \left(H-H_0 \right)^2 + \bar{\kappa} K - \sum_\pm \frac{z_\pm}{A_0} g_\pm \right]
\end{eqnarray}
where $g_\pm(\mathbf{r}) = \int d\theta~\{ \exp [- E_\pm(\theta,\mathbf{r})/(k_B T)] - \exp[- E_F/(k_B T)]\}$, $H(\mathbf{r})=(c_1 + c_2)/2$ is the mean curvature, $H_0$ the spontaneous curvature, $K(\mathbf{r})=c_1 c_2$ the Gaussian curvature.  We have also subtracted off a contribution in $g_\pm(\mathbf{r})$, unphysical in a fixed area ensemble, involving the interaction energy between the proteins and a flat membrane. It is straightforward to generalize Eq. (\ref{eq:energy}) to include multiple species of proteins.

We seek to minimize Eq. (\ref{eq:energy}) with respect to the membrane shape. Before proceeding, it is worth noting that a straightforward minimization will not capture the correlated fluctuations of the membrane and, by proxy, the correlated fluctuations of the proteins themselves. Hence, the proteins interact only through the large-scale equilibrium deformations of the membrane. This restriction can be corrected by computing the fluctuation corrections to Eq. (\ref{eq:energy}).

In order to establish our main points with minimal algebraic complication, we will mostly specialize to the extremely anisotropic case $\lambda_X=\lambda_T = 0$. As in the lamellar experiments, we will assume that proteins can bind to either leaf with equal fugacity, so that $z = z_+ = z_-$. Later, we will reintroduce the remaining couplings to see how this basic picture is modified.
With this simplification, $E_F = e^{-\lambda_L/2}$ and
$E_\pm(\mathbf{r},\theta)/(k_B T) = \lambda_L \left(\xi_1 \cos^2 \theta + \xi_2 \sin^2 \theta \mp 1 \right)/2$
in Eq. (\ref{eq:energy}).

\begin{figure}
\includegraphics[width=3.5in]{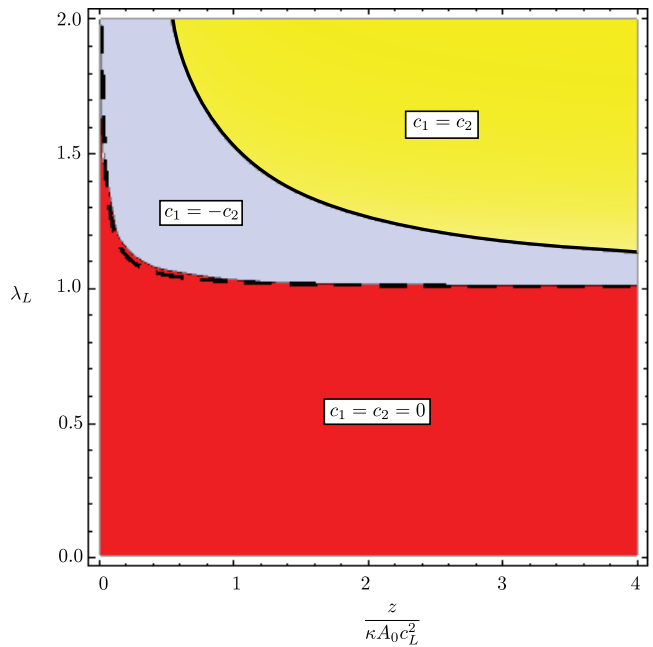}
\caption{\label{fig:fig2} (color online) Longitudinal coupling strength, $\lambda_L$, vs. normalized fugacity, $z/(\kappa A_0 c_L^2)$, phase diagram for a single species with equal binding to both leaves and $\lambda_X=\lambda_T = 0$. We've chosen $\bar{\kappa} = -0.05 \kappa$. The dashed line is the stability line of Eq. (\ref{eq:saddleinstability}) between the flat phase (red/dark gray) and the minimal surface phase (blue/gray). The spherical phase is shown in yellow, with a transition from the $K < 0$ phase that is first order.}
\end{figure}

As a first step toward a topological phase diagram, we ask what combinations of membrane curvatures $\xi_1$ and $\xi_2$ minimize the free energy density.  Fig. \ref{fig:fig2} shows a typical phase diagram as a function of $\lambda_L$ and fugacity $z$. We generically find three morphologies: a flat phase ($\xi_1 = \xi_2 =0$), a saddle phase ($\xi_1 = -\xi_2 \approx 1$), and a spherical phase ($\xi_1 = \xi_2 \approx 1$). When $\bar{\kappa}= 0$, the transition from flat to minimal surface occurs along the line $\lambda_L = 1$; therefore, we use $\bar{\kappa}=-0.05 \kappa$ in Fig. \ref{fig:fig2} to highlight the role of $\bar{\kappa}$.

We can understand the transition from the flat to curved phases in terms of the two ``topological moduli'', $\kappa_+ = \kappa/2 + \bar{\kappa}$ and $\kappa_- = - \bar{\kappa}$ \cite{helfrich, morse}. When $\kappa_+ < 0$, the membrane becomes unstable to topological rearrangements that induce positive Gaussian curvature, for example to spherical vesicles \cite{morse}. Similarly, when $\kappa_- < 0$, the membrane becomes unstable toward the formation of negative Gaussian curvature. By expanding Eq. (\ref{eq:energy}) in powers of the curvature, which is valid only in the flat phase, we find effective moduli are shifted by $\Delta \kappa = \pi (z_+ + z_-)/(A_0 c_L^2) e^{-\lambda_L/2}  3 (1-\lambda_L) \lambda_L$ and $\Delta \bar{\kappa} =\pi (z_++z_-)/(2 A_0 c_L^2) e^{-\lambda_L/2} \lambda_L \left(\lambda_L-1\right)$ \cite{may} so that $\kappa_+ < 0$ when
\begin{equation}\label{eq:sphereinstability}
\frac{z_+ + z_-}{A_0 c_L^2} > 4 \frac{e^{\lambda_L/2}}{\pi \lambda_L (\lambda_L - 1)} \left(\frac{\kappa}{2} + \bar{\kappa} \right),
\end{equation}
and $\kappa_- < 0$ when
\begin{equation}\label{eq:saddleinstability}
\frac{z_+ + z_-}{A_0 c_L^2} > 2 \frac{e^{\lambda_L/2}}{\pi \lambda_L (\lambda_L - 1)} |\bar{\kappa}|.
\end{equation}
Eq. (\ref{eq:saddleinstability}) predicts the boundary between the flat and saddle phases in Fig. \ref{fig:fig2} quite well. If $\bar{\kappa} < - \kappa/3$, the transition from the flat phase proceeds directly to spheres, as can also be seen from Eqs. (\ref{eq:sphereinstability}) and (\ref{eq:saddleinstability}). In contrast, an instability with respect to \textit{continuous} perturbations occurs at a critical protein fugacity $(z_+ + z_-)/(A_0 \kappa c_L^2) > e^{\lambda_L/2}/[3 \pi \lambda_L (\lambda_L-1)]$ above which the effective bending modulus $\kappa +\Delta \kappa < 0$. There is also a shift in the spontaneous curvature of $\Delta H_0 = 2 \pi (z_+ -z_-)/(A_0 c_L) e^{-\lambda_L/2} \lambda_L/(\kappa+\Delta \kappa)$ which vanishes when the fugacities $z_+ = z_- = z$ are balanced.

Why do anisotropic couplings lead to such rich phase diagrams? When $\lambda_X = \lambda_T = 0$, the protein curvature can be accommodated by both $K>0$ and $K< 0$ since they need only find a single direction in which the membrane curvature is commensurate with the protein's intrinsic curvature. However, the orientational entropy of the protein is maximized at umbilics ($\xi_1 = \xi_2$) while the bending energy is minimized at minimal surfaces ($\xi_1 = - \xi_2$). The dependence on the combination $z/\kappa$ arises directly from this competition.
%
\begin{figure}
\includegraphics[width=3.5in]{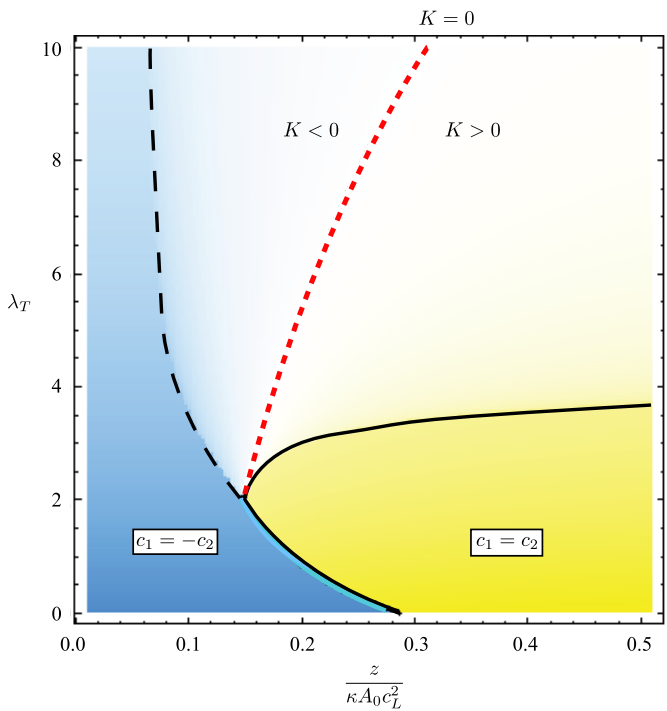}
\caption{\label{fig:fig3} (color online) Transverse coupling strength, $\lambda_T$, vs. renormalized fugacity $z/(\kappa A_0 c_L^2)$ phase diagram ($\bar{\kappa} = 0$) for a single species of protein with $\lambda_X=\lambda_L = 10$ and with equal binding to both bilayer leaves. The solid, black lines indicate a discontinuous transition while the dashed, black line is continuous. There is a change of symmetry across each of these black lines. Cylinders ($K=0$) lie on the red line. The color intensity is indicates the magnitude of the Gaussian curvature.}
\end{figure}

Terms higher than second order in the curvature arising in Eq. (\ref{eq:energy}) are responsible for stabilizing the highly curved phases in Fig. \ref{fig:fig2}. These terms typically compete with potential higher order terms in the Helfrich energy. These Helfrich terms are typically ignored since, in the absence of other physics \cite{fournier3}, they are associated with microscopic length scales, $\ell$. The higher order terms in our model, on the other hand, scale with $(z_+ + z_-) e^{-\lambda_L/2}/(A_0 c_L^2)$, the average protein density on a flat membrane. If we denote $R$ as the average separation between proteins on the membrane, we obtain that the regime of validity of our model, with respect to higher order corrections, $R \ll 1/(t c_L^2)$ for our model to be valid without these additional corrections.

If we introduce the coupling $\lambda_X \ne 0$, the spherical phase increases its range of stability. Even when $\lambda_X = \lambda_L$, however, all three phases remain at higher couplings, though at small couplings ($\lambda_X = \lambda_L \approx 3$) the spherical phase preempts the saddle phase entirely. When $\lambda_T \ne 0$ as well, additional phases can appear. If $\xi_T = 0$, for example, a ``cylindrical'' phase, with $|\xi_1| \ne |\xi_2|$ and $|\xi_2| \ll 1$, is introduced. As an example, a cross-section of the phase diagram along the slice $\lambda_L = \lambda_X = 10$ is plotted in  Fig. \ref{fig:fig3}. The transitions meet at a single critical point at which $\xi_1 = \xi_2 = 0$. For completely isotropic couplings ($\lambda_T=\lambda_X = \lambda_L$), there is a ``cylindrical'' phase (with a small but nonzero $K$) above a critical fugacity and coupling strength. At small but isotropic couplings, however, protein entropy dominates so that only the flat and spherical phase remain.

Additional complications arise when we make the transition from local phase diagrams, such as those in Figs. \ref{fig:fig2} and \ref{fig:fig3}, to equilibrium membrane morphologies. We focus on the case that $\lambda_X = \lambda_T = 0$ and $\bar{\kappa}=0$. In the case of the spherical phase, the lowest energy state is one of monodisperse spherical vesicles of radius $c_L^{-1}$. In the saddle phase, however, the lowest energy state occurs when $\xi_1 = -\xi_2 \approx 1$. This is impossible in a three dimensional Euclidean space -- the membrane energy is frustrated by the constraints of geometry. As a prototype of a topological transition driven by proteins, we consider the lamellar to bicontinuous phase transition observed experimentally in the presence of certain curvature-inducing peptides \cite{wong2}. Generically, lamellar phases that have a tendancy toward negative Gaussian curvature can display a variety of complex morphologies \cite{didonna}; in many cases, however, these complex layered structures can be built from a superposition of simple defects \cite{schnerk}. In fact, screw dislocations are the building blocks of more complex minimal surfaces \cite{colding}.

Here, we will consider the free energy of inserting a single screw dislocations into the lamellar order. Once it becomes energetically favorable to insert dislocations, we should expect them to proliferate in the ground state. Since screw dislocations of either sign will be degenerate, we should expect the formation of a globally achiral structure such as that explored in Ref. \cite{schnerk}.
\begin{figure}
\includegraphics[width=3.5in]{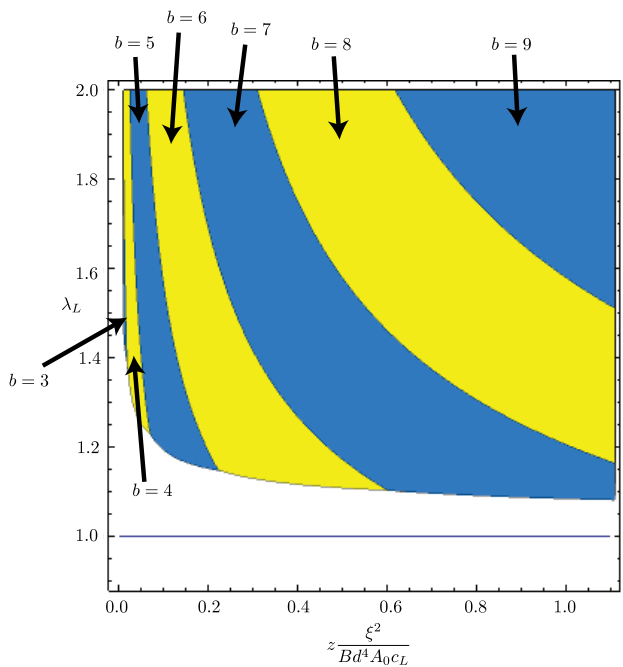}
\caption{\label{fig:fig4} (color online) Longitudinal coupling strength, $\lambda_L$, vs. renormalized fugacity $z \xi^2/(B d^4 A_0 c_L)$ critical line above which helicoids are energetically favorable for $\bar{\kappa}=0$. The flat phase is stable in the region $\lambda_L < 1$. Color (shading) is added as a guide to the eye.}
\end{figure}
In terms of coordinates $(x,y)$ on a flat reference layer, a screw dislocation can be described by the multivalued height function of a helicoid, $h(x,y) = b \tan^{-1} (y/x)/(2 \pi)$, where $b = n d$ is the Burgers scalar, $n$ an integer, and $d$ the layer spacing. In the absence of proteins, this is an extremum of both the bending and compression energies \cite{degennes}; since we are only after the stability of the lamellar phase, we will assume the same height function with proteins present. Screw dislocations have a large elastic contribution to their line tension, arising from the deviation from the equilibrium layer spacing near the core, of the form $\tau = B b^4/(256 \pi^4 \xi^2)$, where $B$ is the bulk modulus and $\xi$ is a microscopic core size \cite{degennes}. Due to the divergence as $\xi \rightarrow 0$, we neglect an additional core energy in comparison to the elastic line tension.

Since $H=0$, this line tension must be balanced with the contribution from Eq. (\ref{eq:energy}), which we evaluate numerically. We find that a screw dislocation becomes energetically favorable when compared to the flat phase above the critical line shown in Fig. \ref{fig:fig4}. Increasing fugacity prefers an increasing burgers scalar, with both chiralities degenerate. This critical line occurs at larger $\lambda_L$ than predicted by the local phase diagram. The transition also depends strongly on the bulk modulus of the lamellar phase, though we note that the layer spacing is typically set by $d \sim \sqrt{(\kappa/d)/B}$ \cite{degennes}. If we use $\xi \sim d$ for the core size we obtain that $z \xi^2/(B d^4 A_0 c_L) \sim z c_L d/(\kappa A_0 c_L^2)$, so the horizontal axis of Fig. \ref{fig:fig4} is roughly comparable to that of our local phase diagrams when $c_L \sim 1/d$.

Finally, the dependence of our phase diagrams on membrane rigidity imply that the induced morphology also depends on lipid composition, since lipids with intrinsic curvature can modify the membrane rigidity \cite{may}. For an anisotropic lipid with $\lambda_L < 1$, for example, our results indicate that $\Delta \kappa > 0$, suggesting that these intrinsically-curved lipids can drive a membrane coated with proteins from a spherical to saddle phase. Should curvature-inducing proteins take advantage of our mechanism for sculpting membranes, tuning the membrane moduli by adding cosurfacants or changing lipid mixtures will induce a corresponding transition in membrane Gaussian curvature.

In summary, we have studied topological transitions in membranes induced by anisotropic, curvature-inducing proteins. We predict a surprising variety of morphologies having both positive and negative Gaussian curvatures depending on the protein densities and membrane moduli. In a regime of negative Gaussian curvature, this induces a transition from a lamellar phase to one with screw dislocations of both chiralities in the ground state.

This work was funded by the National Science Foundation through DMR-0846582.

\end{document}